# Elliptically polarized terahertz radiation from a chiral oxide

R. Takeda[1], N. Kida[1,*], M. Sotome[1], and H. Okamoto[1]

[1]*Department of Advanced Materials Science, The University of Tokyo, 5-1-5 Kashiwa-no-ha, Chiba 277-8561, Japan*

**Abstract**

Polarization control of terahertz wave is a challenging subject in terahertz science and technology. Here, we report a simple method to control polarization state of the terahertz wave in terahertz generation process. At room temperature, terahertz radiation from a noncentrosymmetric and chiral oxide, sillenite $Bi_{12}GeO_{20}$, is observed by the irradiation of linearly polarized femtosecond laser pulses at 800 nm. The polarization state of the emitted terahertz wave is found to be elliptic with an ellipticity of ~0.37±0.10. Furthermore, the ellipticity was altered to a nearly zero (~0.01±0.01) by changing the polarization of the incident linearly polarized femtosecond laser pulses. Such a terahertz radiation characteristic is attributable to variation of the polarization state of the emitted terahertz waves, which is induced by retardation due to the velocity mismatch between the incident femtosecond laser pulse and generated terahertz wave and by the polarization tilting due to the optical activity at 800 nm.

* E-mail: kida@k.u-tokyo.ac.jp

# I. Introduction

Recent progress of terahertz science and technology has opened new applications of terahertz waves for various kinds of existing tools such as spectroscopy, imaging, and sensing [1]. In order to improve the performance of these terahertz applications, control of polarization state of the terahertz wave is necessary. In this context, two approaches have been proposed. One approach is the use of terahertz optical components such as polarizers, waveplates, and modulators, which are fabricated by, for examples, multiple-quantum-wells [2-4], metamaterials [5-8], carbon nanotubes [9, 10], and birefringent media [11, 12]; elliptically or circularly polarized terahertz wave can be obtained passing through the terahertz waveplates. Other approach is the direct control of the terahertz waves in terahertz generation process. By the irradiation of a pair of femtosecond laser pulses to noncentrosymmetric media such as ZnTe, two orthogonally polarized terahertz waves can be generated. By combining the generated terahertz waves with an appropriate interval of two femtosecond laser pulses [13-18] or an arbitrary shaped femtosecond laser pulses [19], the polarization-controlled terahertz radiation was obtained.

Here, we propose that a noncentrosymmetric chiral media can act as the polarization-dependent terahertz wave emitter with a controllable ellipticity. Chirality –its mirror image cannot be superimposed by any rotation and reflection– emerges in various fields of physics, chemistry, and biology. The unique optical property of the chiral media is the optical activity [20], in which the polarization of light rotates when the light propagates in the medium. This phenomenon can be understood by the difference of the optical response between right- and left-handed circularly polarized light. Recent experiments have revealed the presence of the optical activity even in two-dimensional artificial media [21] and magnetic compound [22].

Due to the optical activity in the chiral media, the electric field of the incident linearly polarized femtosecond laser pulse rotates during the propagation. As a result, terahertz waves with different polarizations are generated. Provided that the velocity mismatch between the femtosecond laser pulse and the terahertz waves is significant in the medium, the terahertz waves with their polarization tilted to each other superimpose with various phase shifts. This would result in the emission of elliptically polarized terahertz waves outside the medium. However, emission of elliptically polarized terahertz waves in terahertz radiation process has never been investigated so far in noncentrosymmetric and chiral media.

In this work, we focus on a noncentrosymmetric and chiral oxide, sillenite $Bi_{12}GeO_{20}$. Crystal structure of $Bi_{12}GeO_{20}$ is cubic with a point group of *T* or 23, as illustrated in Fig. 1(a). At room temperature, $Bi_{12}GeO_{20}$ lacks the space inversion symmetry. Thus, it shows the piezoelectric effect [23] and nonlinear optical effect with a relatively large electro optic coefficient of ~3.29 pm/V at 850 nm [24, 25]. Furthermore, it was reported that $Bi_{12}GeO_{20}$ exhibits the optical activity in the transparent frequency region; the rotation of the transmitted light [24], characterized by the rotatory power $\rho$, was ~10°/mm at ~800 nm [23].

In the present paper, we report on the first observation of the emission of terahertz waves from $Bi_{12}GeO_{20}$ by the irradiation of a femtosecond laser pulse at room temperature. The origin of the terahertz radiation was optical rectification by second-order nonlinear optical process. In the power spectrum of the terahertz radiation, we observed two sharp peak structures at ~1.42 THz and ~1.64 THz and an intensity fringe pattern with a period of ~0.08 THz below 0.4 THz. By taking into account the effective generation length for the terahertz radiation, we reproduced not only two sharp peak structures but also the intensity fringe pattern. Notably, we found that the emitted terahertz wave becomes elliptic with a maximum ellipticity of ~0.37±0.10. Furthermore, the ellipticity depends on the polarization state of an incident linearly polarized femtosecond laser pulse; the ellipticity was changed from ~0.37±0.10 (elliptically polarized) to ~0.01±0.01 (linearly polarized). On the basis of comprehensive optical measurements in terahertz and visible frequency regions, we discuss the observed elliptically controllable terahertz radiation from Bi$_{12}$GeO$_{20}$ in terms of the phase-matching condition and optical activity at 800 nm.

## II. Second-order nonlinearly optical susceptibility tensor

Here, we explain the terahertz radiation mechanism by second-order nonlinear optical effect in a noncentrosymmetric media. When a noncentrosymmetric media is irradiated by a femtosecond laser with a finite spectral width, a transient electric polarization $P$ can be induced in the picosecond time scale via difference frequency mixing. In turn, this picosecond modulation results in the emission of terahertz waves into free space. Such optical rectification can be represented by $P = \varepsilon_0 \chi^{(2)}(\omega-\omega=0)E^{\omega}E^{\omega}$, where $\varepsilon_0$ is the dielectric constant of vacuum and $\chi^{(2)}$ is the second-order nonlinear optical susceptibility [29]. This is recognized as the general mechanism of the terahertz radiation in various kinds of noncentrosymmetric media such as ZnTe [29]. Nonzero tensor components of $\chi^{(2)}$ of Bi$_{12}$GeO$_{20}$, are $\chi_{xyz} = \chi_{yzx} = \chi_{zxy} = \chi_{xzy} = \chi_{yxz} = \chi_{zyx}$ [29]. The contraction of $\chi^{(2)}$ to the nonlinear $d$ tensor is $d_{14}$. $P$ of a (110)-oriented crystal can be expressed as,

$$\begin{bmatrix} P_{1\bar{1}0} \\ P_{110} \\ P_{001} \end{bmatrix} = \varepsilon_0 d_{14} E_0^2 \begin{bmatrix} \sin 2\theta \\ 0 \\ \cos^2\theta \end{bmatrix}, \qquad (1)$$

where $E_0$ is the electric field of a femtosecond laser pulse. We defined the angle $\phi$ as the angle of the electric field of the terahertz wave relative to the $[1\bar{1}0]$ direction of the crystal; $\phi$ is given by

$$\phi = \operatorname{atan}\left(\frac{\cos^2\theta}{\sin 2\theta}\right). \qquad (2)$$

## III. Experimental details

We used a commercially available 1-mm-thick (110)-oriented single crystal of $Bi_{12}GeO_{20}$. In the terahertz frequency region, we measured the reflectance $R$ and transmittance $T$ spectra by using Fourier-transformed infrared (FT-IR) spectroscopy (20−600 cm$^{-1}$) and terahertz time-domains spectroscopy, respectively. For the terahertz radiation experiments, we adopted the standard photoconducting sampling technique using a low-temperature-grown GaAs detector. We used a femtosecond laser pulse (the center wavelength of 800 nm, the repetition rate of 80 MHz, and the pulse width of 100 fs) delivered from a mode-locked Ti:Sapphire laser. A linearly polarized pump pulse was focused with a spot diameter of 25 μm and the laser power was 153 μJ/cm$^2$ per pulse. We defined the laboratory coordinate as $X$-(horizontal) axis and $Y$-(vertical) axis, as illustrated in Fig. 1(b). The wire-grid polarizer (WG1) was inserted in front of the detector so as to detect only the $X$-axis component of the terahertz wave. All the measurements were performed at room temperature.

## IV. Results and discussion
## A. Optical spectra

First, we show the optical spectra of $Bi_{12}GeO_{20}$ in the terahertz frequency region. The measured $R$ and $T$ spectra are shown by circles and squares in Fig. 1(d), respectively. We performed the dispersion analysis of the $R$ spectrum up to ~10 THz by assuming twelve Lorentz oscillators, the result of which below 2.5 THz is indicated by the solid line in Fig. 1(d); the estimation procedure is detailed in Appendix A. Figures 1(e) and (f) show the calculated refractive index $n$ and absorption coefficient $\alpha$ spectra, respectively. Both spectra are in good agreement with the $n$ and $\alpha$ spectra (symbols) in the frequency range of 0.3–1 THz, which were directly obtained by terahertz time-domain spectroscopy in transmission geometry (see Appendix A). There are two dispersions around 1.33 THz and 1.51 THz in the $n$ spectrum [Fig. 1(e)]. Accordingly, the sharp peak structures show up at ~1.37 THz and ~1.54 THz in the $\alpha$ spectrum [Fig. 1(f)]. According to Raman scattering experiments (Refs. 23 and 27), Raman active transverse optical (TO) $F$ modes, assigned to the external modes of Bi atoms, were observed at ~1.37 THz and ~1.54 THz. Since the $F$ modes are also infrared active, the observed peak structures in the $\alpha$ spectrum are also attributable to these modes.

**B. Terahertz radiation**

Next, we show the results of terahertz radiation experiments. The penetration depth of a femtosecond laser pulse was evaluated to be ~3.33 mm, which exceeds the thickness of the sample (1 mm) (see Appendix A). In this experiment, the electric field of a femtosecond laser pulse was set parallel to the $X$-axis. The angle $\varphi$ was defined as the angle of the $[1\bar{1}0]$ direction of the crystal relative to the $X$-axis, as illustrated in Fig. 1(b). Upon irradiation of femtosecond laser pulses, we found that the terahertz wave is radiated from $Bi_{12}GeO_{20}$ into free space. The measured waveform with $\varphi = 35°$ is shown in Fig. 1(c). It consists of temporal oscillation components. Figure 1(i) shows the Fourier transformed spectrum of the waveform, which contains the frequency component up to ~2.5 THz. Two dip structures indicated by arrows appear at ~1.37 THz and ~1.54 THz; their frequency positions just correspond to the peak positions of the TO $F$ modes, as seen in the $\alpha$ spectrum [Fig. 1(f)]. In addition, an intensity fringe pattern with a period of ~0.08 THz is discerned below 1.4 THz, the origin of which will be discussed later. In the same experimental setup, we measured the terahertz waveform in a 0.5-mm-thick (110)-oriented ZnTe crystal, which is known as a typical terahertz wave emitter. The maximum amplitude of the terahertz electric field at 0 ps in $Bi_{12}GeO_{20}$ was about 1/80 of that in ZnTe.

## C. Terahertz radiation mechanism

In order to reveal the terahertz radiation mechanism in $Bi_{12}GeO_{20}$, we measured the azimuth angle dependence of the amplitude of the terahertz electric field at 0 ps ($E_{THz}$). In this experiment, $\varphi$ was fixed to be 0°. The electric field of femtosecond laser pulses was rotated with the angle $\theta$ by a half-wave plate, as illustrated in Fig. 2(a). $\theta$ was defined as the angle of the electric field of femtosecond laser pulses relative to the $X$-axis. In order to detect the $X$- and $Y$-axes components of $E_{THz}$ in the same experimental condition, we performed the vector analysis of the polarization. In addition to the WG1, we inserted another wire-grid polarizer (WG2). The angle of WG2 was set to be +45° or -45° with respect to the $Y$-axis. In this setup, the components of $E_{THz}$ along the $X$- and $Y$-axes, respectively, can be expressed as

$$E_X \propto (E_{+45°} + E_{-45°}), \qquad (3)$$

$$E_Y \propto (E_{+45°} - E_{-45°}), \qquad (4)$$

where $E_{+45°}$ and $E_{-45°}$ are $E_{THz}$ obtained in the angle of WG2 at +45° and -45°, respectively. By the vector composition of $E_{+45°}$ and $E_{-45°}$ [Fig. 2(b)], we estimated both $E_X$ and $E_Y$, the results of which are shown by circles in Figs. 2(c) and (d), respectively. We also show in Fig. 2(e) the rotation angle $\phi$ (circles) of the terahertz wave, as illustrated in Fig. 2(a). $\phi$ is given by $\phi = \mathrm{atan}(E_Y / E_X)$. As can be seen, $E_X$, $E_Y$, and $\phi$ depend on $\theta$. $E_X$ reaches the maxima at ~70° and its sign was reversed by 90°, while $E_Y$ shows the maxima at ~10° and becomes zero at ~100°. As a result, $\phi$ changes from -90° to 90°. In order to explain their $\theta$ dependences, we consider optical rectification by second-order nonlinear optical effect [see Eqs. (1) and (2)]. The blue lines in Figs. 2(c) and 2(d) indicate the $\theta$ dependence of the calculated $E_X$ and $E_Y$ using Eq. (1) in Sect. II, respectively. As can be seen, the $\theta$ dependence of the measured $E_X$ and $E_Y$ (symbols) are qualitatively reproduced, except for the shift of $\theta$ ~24.6° for $E_X$ and $\theta$ ~8.7° for $E_Y$. We reproduced the $\theta$ dependence of the measured $\phi$ by Eq. (2) in Sect. II with fitting parameters used for $E_X$ and $E_Y$, as indicated by the solid line in Fig. 2(e). This indicates that the terahertz radiation in $Bi_{12}GeO_{20}$ is due to optical rectification.

Next, we discuss the reason why the shift of $\theta$ ~24.6° for $E_X$ and $\theta$ ~8.7° for $E_Y$ appears. Since $Bi_{12}GeO_{20}$ is known to show the optical activity at 800 nm [23], we measured the rotation of the electric field of femtosecond laser pulses passing through the same sample. $\rho$ was estimated to be ~12°/mm and was independent of $\theta$, being in good agreement with the previous result (~10°/mm) [23]. The measured $\rho$ is roughly consistent with the observed shift of $\theta$ ~8.7° for $E_Y$. Thus, we calculated the $\theta$ dependence of $E_Y$ by taking into account the shift of $\theta$ (~8.7°); the measured $\theta$ dependence of $E_Y$ is almost reproduced, as indicated by the black line in Fig. 2(d). On the contrary, the large shift of the $E_X$ (~24.6°) cannot be explained by the optical activity at 800 nm alone. Although its origin is unclear, the observed shift of $\theta$

~24.6° for $E_X$ implies the presence of polarization-dependent effects such as photo-refractive effect in the terahertz frequency region. In order to reveal the observed shift for $E_X$, further measurements of optical anisotropy upon irradiation of the femtosecond laser pulse are needed. In this paper, we focus on terahertz radiation characteristics of $Bi_{12}GeO_{20}$.

In the terahertz radiation process, phase-matching condition and attenuation of the terahertz waves play crucial roles. We estimated the coherence length $l_c$ [30] and the effective generation length $l_c^{eff}$ [31] for the terahertz radiation in optical rectification, as detailed in Appendix B. The group refractive index $n_g$ at 800 nm was estimated to be ~2.79, which is indicated by the horizontal line in Fig. 1(e). $l_c$ as a function of frequency is shown by the purple line in Fig. 1(g), in which two peak structures at ~1.4 THz and ~1.6 THz are discerned. The increases of $l_c$ at these frequencies are due to the presence of two sharp dispersions in the $n$ spectrum associated with TO $F$ modes [Fig. 1(e)]. In order to include the effect of the strong absorptions of these modes [Fig. 1(f)], we calculated $l_c^{eff}$ as a function of frequency, which is shown by the red line in Fig. 1(g). The length for the terahertz radiation totally decreases. Furthermore, intensity fringe pattern with a period of ~0.08 THz appears below 1 THz due to large mismatch of $|n-n_g|$ (~4.1) [see Eq. (8) in Appendix B]. Since $l_c^{eff}$ (~5−20 μm) is shorter than the thickness of the sample (1 mm), the terahertz wave generates at the back region of the sample, as illustrated in Figs. 3(l). We further considered the instrumental function of the detection system; our calculation procedure is detailed in Ref. 26 for the study of α-TeO$_2$. The calculated spectrum is shown in Fig. 1(h), which reproduces not only two sharp peak structures at ~1.42 THz and ~1.64 THz but also the intensity fringe pattern with a period of ~0.08 THz below 1.4 THz. Thus, we conclude that the spectral shape can be determined by phase-matching condition for the terahertz radiation.

## D. Elliptically polarized terahertz radiation

In order to reveal the effect of the optical activity on the terahertz radiation characteristics in a chiral $Bi_{12}GeO_{20}$, we measured $E_X$ and $E_Y$ as a function of delay time $t$. The measured $E_X(t)$ and $E_Y(t)$ with $\theta =114°$ are shown by blue and red lines in Fig. 3(f), respectively. A black line represents the three-dimensional trajectory of the terahertz wave, which was obtained by the vector composition of $E_X(t)$ and $E_Y(t)$. For convenient, the projection of the terahertz wave onto $E_X$-$E_Y$ plane is shown by the green line. As can be seen, the elliptically polarized terahertz radiation was observed. Figure 3(d) shows the one round of data (circles), which can be reproduced by the ideal ellipse (the solid line) with an ellipticity $\eta$ of ~0.37±0.10. We measured the $\theta$ dependence of the polarization state. Figures 2(g) and 3(b) show the trajectory of the terahertz wave and the one round of data, respectively, with $\theta =$ 42°. A linearly polarized terahertz radiation with $\eta$ of ~0.01±0.01 was observed. To clearly see the change of the polarization state of the terahertz wave, we show in Figs. 3(a)-(h) the selected one round of data at various $\theta$; the number in the figures indicates the value of $\theta$. Figure 3(i) summarizes the $\theta$ dependence of the measured $\eta$. $\eta$ reaches the maxima at $\theta =$ 114° and $\theta =$ 295°. In addition, peak structures are observed at $\theta =$ 10°, $\theta =$ 70°, $\theta =$ 190°, and $\theta =$ 250°.

Here, we discuss why polarization state of the emitted terahertz wave becomes elliptic by taking account the optical activity at 800 nm and the large mismatch of $n$-$n_g$ (~4.1) [Fig. 1(e)]. Since the estimated $l_c$ is ~10–80 μm [Fig. 1(g)], the terahertz wave generates at the back region of the sample. For simplify, we consider "A" and "B" areas within the area $L$, which are schematically shown in Figs. 3(j) and 3(k), respectively. At "A" area, terahertz wave ($E_{THz}^A$) is generated by the irradiation of a femtosecond laser pulse [Fig. 3(j)]. The generated $E_{THz}^A$ proceeds with the phase velocity of $c/n$ ($c$ is the velocity of light) and reaches at "B" area. On the other hand, a femtosecond laser pulse proceeds with the group velocity of $c/n_g$ and reaches at "B" area. As a result, another terahertz wave ($E_{THz}^B$) generated at "B" area has retardation to $E_{THz}^A$. Furthermore, the polarization state of the femtosecond laser pulse rotates at "B" area, compared to "A" area, due to the optical activity. Consequently, the retardation and polarization tilting between $E_{THz}^A$ and $E_{THz}^B$ occur, resulting in emission of elliptically polarized terahertz waves. In order to confirm the discussion above, we performed a simple model calculation. We divided the area within $L$ ~80 μm into $N$ areas, which is labeled by "$i$", as illustrated in Fig. 3(l). The $[1\bar{1}0]$ and [001] components of the terahertz wave, respectively, can be simply expressed as

$$E_{[1\bar{1}0]} = \sum_{i=1}^{N} \sin 2(\theta - \rho'(d - L\frac{i}{N}))I(t')\exp(-\alpha L\frac{i}{N}), \qquad (5)$$

$$E_{[001]} = \sum_{i=1}^{N} \cos^2(\theta - \rho''(d - L\frac{i}{N}))I(t')\exp(-\alpha L\frac{i}{N}), \qquad (6)$$

where $I(t')$ is the ideal terahertz waveform, given by $I(t') \propto \exp(-2t'^2/\tau^2)$ with $\tau = 0.74$ (ps). By taking into account the phase delay between a femtosecond laser pulse and the generated terahertz wave, $t'$ is given by $t' = t + i\frac{L}{N}\frac{n-n_g}{c}$. In this calculation, we used the observed shift of $\theta$ as $\rho'$ (~24.6°) [Fig. 2(c)] and $\rho''$ (~8.7°) [Fig. 2(d)]. $N$ was set to be 50. We included the effect of the absorption during the propagation of the terahertz wave, which is characterized by $\alpha$ at 0.5 THz (~20 cm$^{-1}$). We calculated the polarization of the terahertz waveform and estimated $\eta$ around 0 ps as a function of $\theta$, the result of which is shown by the solid line in Fig. 3(i). Despite simple calculation, the calculated curve reproduces the trend of the measured $\eta$. The remnant discrepancy is perhaps due to the assumption of the dispersionless of $l_c$.

## V. Summary

In summary, we observed terahertz radiation at room temperature in a noncentrosymmetric and chiral oxide, sillenite $Bi_{12}GeO_{20}$, by the irradiation of a femtosecond laser pulse, which was successfully explained by the optical rectification process. We found that the polarization state of the generated terahertz wave becomes elliptic. By changing the polarization state of an incident femtosecond laser pulse, we can control the ellipticity from ~0.37±0.10 to a nearly zero (~0.01±0.01). Our study presents a new method to generate terahertz radiation with an elliptical polarization, which is characteristic to a noncentrosymmetric and chiral material and has an important implication for the applications of terahertz waves.

## Acknowledgement

We thank A. Doi, J. Fujioka, and Y. Tokura for their support in far-infrared reflectance measurements. This work was partly supported by a Grant-in-Aid by MEXT (No. 25247049, No. 25247058, and No. 25-3372). M. S. was supported by Japan Society for the Promotion of Science (JSPS) through Program for Leading Graduate Schools (MERIT) and JSPS Research Fellowships for Young Scientists.

# Appendix

## A. Estimation of complex optical constants

We measured the reflectance $R$ and transmittance $T$ spectra in the energy range of 0.5–5 eV by the grating monochromator. Figure 4(d) shows the measured $R$ and $T$ spectra. A sharp decrease of $T$ and the gradual increase of $R$ are due to the presence of the charge-transfer gap around 2.2 eV. We show in Fig. 4(f) the absorption coefficient $\alpha$ spectrum, which was estimated from $R$ and $T$ data by using the relationship of $\alpha = -\frac{1}{d}\ln\left(\frac{T}{(1-R)^2}\right)$, where $d$ is the thickness of the sample. $Bi_{12}GeO_{20}$ has a transparent window below 2 eV. $\alpha$ at 800 nm (the wavelength of a femtosecond laser pulse used for the terahertz radiation experiments) was estimated to be ~3 cm$^{-1}$, which corresponds to the penetration depth of ~3.33 mm. It exceeds the thickness of the sample (1 mm). By performing the Kramers-Kronig transformation using the measured $R$ and $T$ spectra, we estimated the refractive index $n$ spectrum, as shown by the blue line in Fig. 4(e). As indicated by the broken line in Fig. 4(e), we reproduced the $n$ spectrum by using the Sellmeier relationship, given by $n = \sqrt{1 + \frac{S_0 \lambda_0^2}{1-(\lambda_0/\lambda)^2}}$ with $S_0$ = 5.920 m$^{-2}$ and $\lambda_0$ = 278 nm, where $\lambda$ is the wavelength. The calculated optical group refractive index $n_g$ spectrum is shown by the green line in Fig. 4(e). $n_g$ is given by $\left|n - \lambda \frac{dn}{d\lambda}\right|$. $n_g$ at 800 nm was estimated to be ~2.79, which is indicated by the horizontal line in Fig. 4(b).

Figure 4(a) shows the optical spectra in the terahertz frequency region up to 10 THz. $R$ (circles) and $T$ (squares) spectra were measured by Fourier-transformed infrared (FT-IR) spectroscopy and terahertz time-domain spectroscopy, respectively. $R$ and $T$ spectra below 2.5 THz are shown in Fig. 1(d). In order to estimate the $n$ and $\alpha$ spectra up to 10 THz, we performed the dispersion analysis by assuming a Lorentz oscillator with a damping rate $\gamma$ and the oscillator strength $f$. The measured $R$ spectrum was able to be reproduced by Lorentz oscillators with parameters listed in Table I. The detailed procedures of our dispersion analysis were reported in Refs. 18 and 26. Figures 4(b) and (c) show the estimated $n$ and $\alpha$ spectra (solid lines), respectively. $n$ and $\alpha$ spectra indicated by squares were directly obtained by the measured transmittance and phase spectra. The detailed our estimation procedure was reported in Ref. 26.

## B. Estimation of coherence length and effective generation length for terahertz radiation

We estimated the coherence length $l_c$ for the terahertz radiation [30]. $l_c$ is given by

$$l_\text{c} = \frac{\pi c}{\omega_\text{THz}|n_\text{g} - n_\text{THz}|}, \tag{B1}$$

where $c$ is the velocity of light. The estimated $l_\text{c}$ using values of $n$ in the terahertz frequency region [Fig. 4(b)] and $n_\text{g}$ at 800 nm (2.79) is shown by the purple line in Fig. 1(g). Strong absorptions, which were assigned to the TO $F$ modes, exist at ~1.37 THz and ~1.54 THz, as seen in the $\alpha$ spectrum shown in Fig. 1(f). Thus, we calculated the effective generation length for the terahertz radiation $l_\text{c}^\text{eff}$ [31] by taking into account the contributions of TO $F$ modes. $l_\text{c}^\text{eff}$ is expressed as

$$l_\text{c}^\text{eff} = \left( \frac{1 + \exp(-\alpha d) - 2\exp\left(-\frac{\alpha}{2}d\right)\cos\left(\frac{\omega}{c}|n_\text{THz} - n_\text{g}|d\right)}{\left(\frac{\alpha}{2}\right)^2 + \left(\frac{\omega}{c}\right)^2 (n_\text{THz} - n_\text{g})^2} \right)^{\frac{1}{2}}, \tag{B2}$$

The calculated $l_\text{c}^\text{eff}$ is shown by the red line in Fig. 1(g).

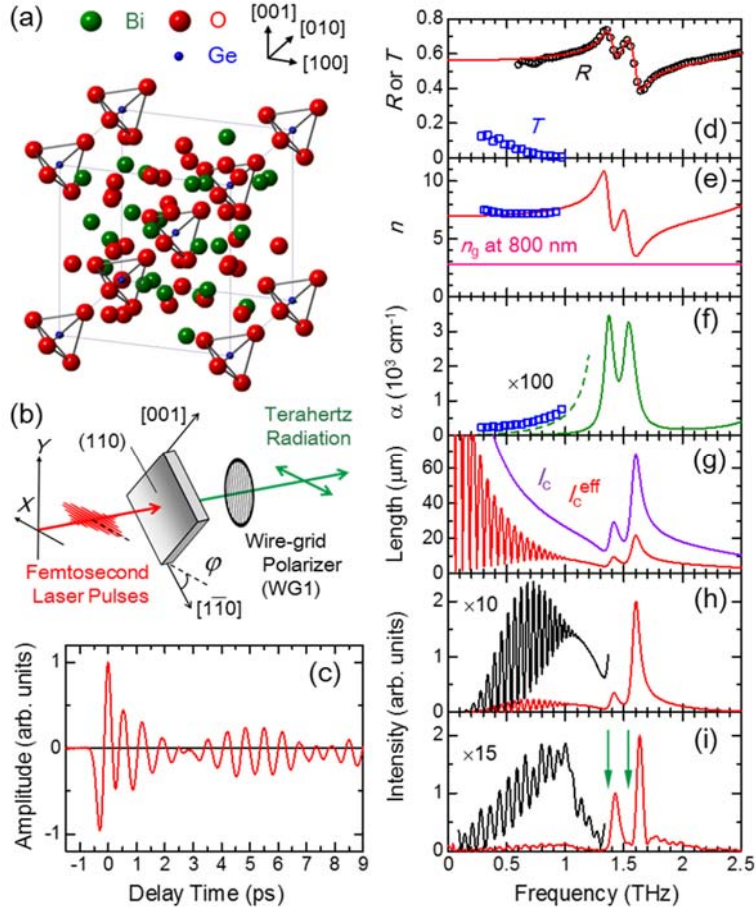

FIG. 1: (Color online) Schematic illustrations of (a) crystal structure of $Bi_{12}GeO_{20}$ and (b) experimental setup. (c) Measured terahertz waveform with $\varphi = 35°$. (d) Reflectance $R$ (circles) and transmittance $T$ (squares) spectra in the terahertz frequency region. The solid line in $R$ indicates the fitting result of the dispersion analysis (see Appendix A). (e) Refractive index $n$ and (f) absorption coefficient $\alpha$ spectra. The horizontal line indicates the value of the group refractive index at 800 nm (see Appendix A). (g) Coherence length $l_c$ and effective generation length $l_c^{eff}$ for the terahertz radiation as a function of frequency. (h) Calculated and (i) measured power spectra of terahertz radiation. The arrows indicate the peak positions of the two $F$ modes as seen in the $\alpha$ spectrum [(f)].

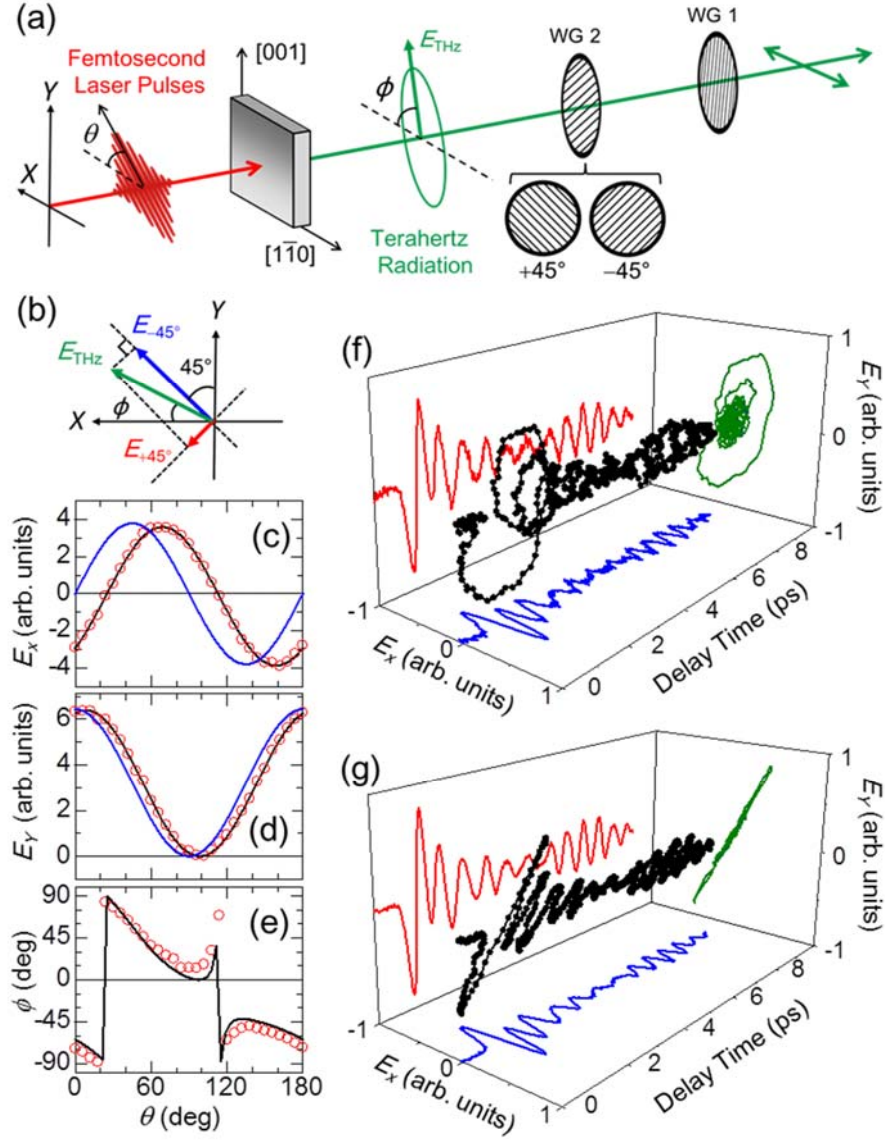

FIG. 2: (a) and (b) Schematic illustrations of experimental setup for the vector analysis. (c) Horizontal $E_X$ and (d) vertical $E_Y$ components (circles) of the terahertz electric field $E_{THz}$ as a function of the azimuth angle $\theta$. (e) $\theta$ dependence of the rotation angle $\phi$. The blue and black lines indicate the fitting results (see text). Measured trajectory of the terahertz wave with (f) $\theta = 114°$ and (g) $\theta = 42°$.

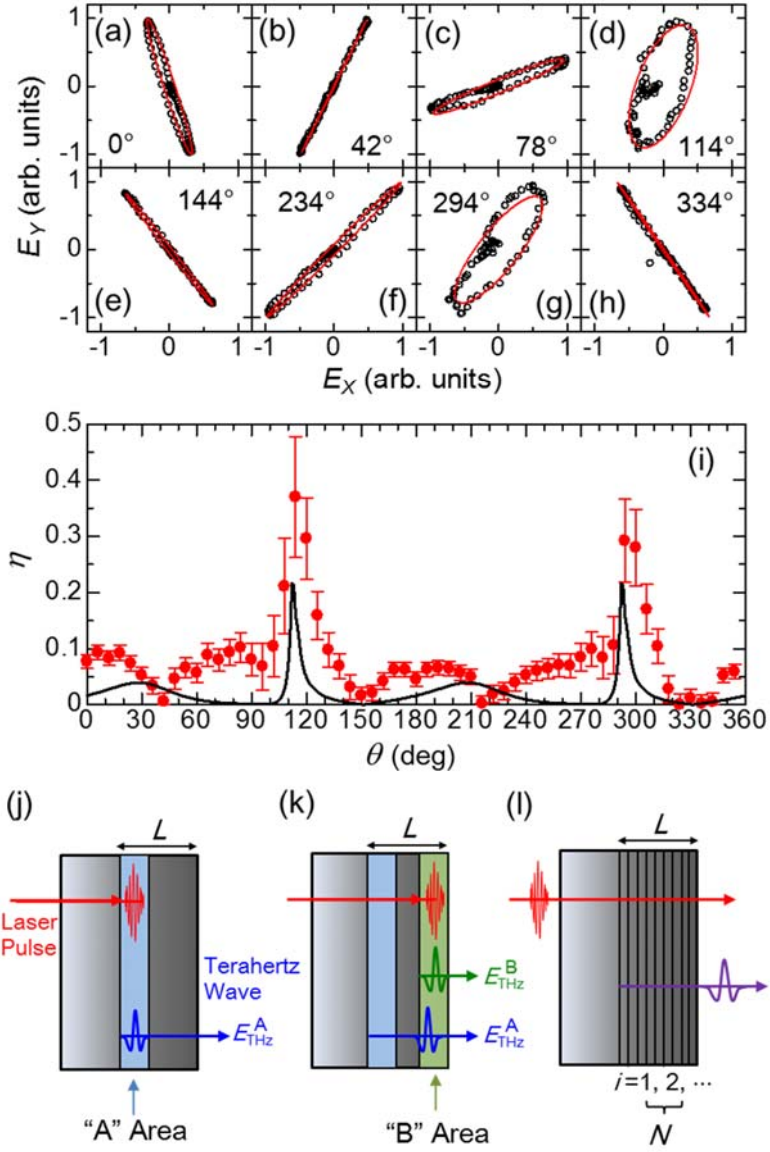

FIG. 3: (a)-(h) Projection of the trajectory of the terahertz wave (circles) onto $E_X$-$E_Y$ plane. The numbers indicate the value of the angle $\theta$. Solid lines indicate the ideal curves of the ellipse. (i) Ellipticity $\eta$ as a function of $\theta$. The solid line indicates the calculated curve (see text). The schematic illustrations of emergence of the retardation and polarization tilting between terahertz waves generated at "A" and "B" areas [(j) and (k)] and (l) our simple model.

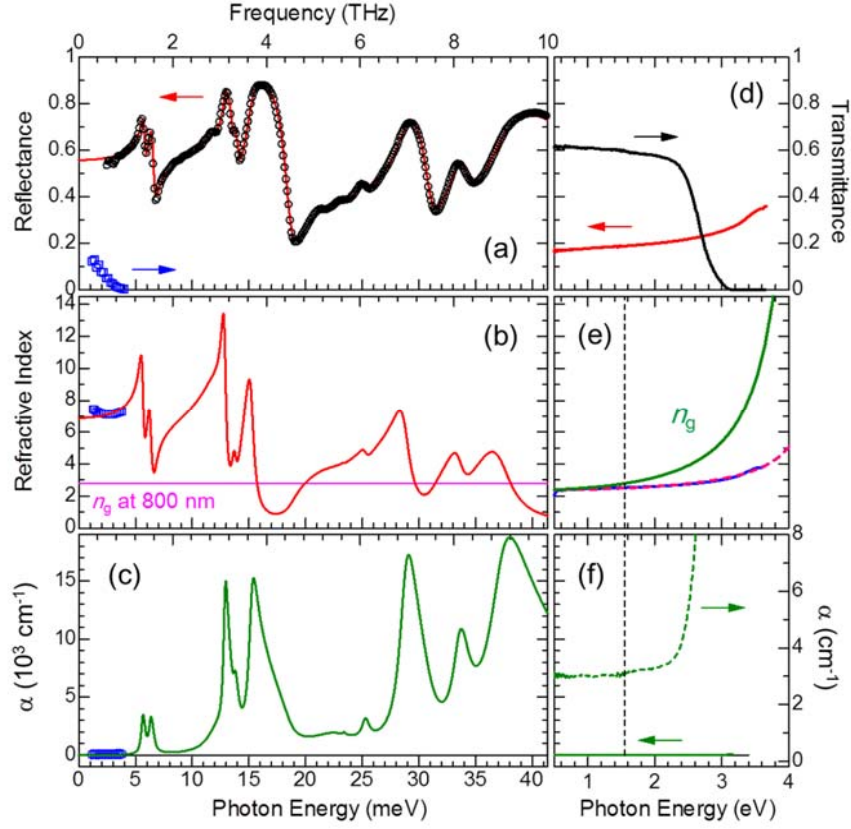

FIG. 4: (a) Reflectance $R$ and transmittance $T$ spectra measured by Fourier-transformed infrared spectroscopy and terahertz time-domain spectroscopy, respectively. The solid line is the result of least-square fit by the dispersion analysis with parameters listed in Table I. The estimated refractive index $n$ [(b)] and the absorption coefficient $\alpha$ [(c)] spectra. (d) $R$ and $T$ spectra measured by the grating spectroscopy. (e) The estimated optical refractive index $n$ spectrum (the blue line) by the Kramers-Kronig transformation. The broken line indicates the fitting result by using Sellmeier relationship. The optical group refractive index $n_g$ spectrum is shown by the green line. (f) $\alpha$ spectrum obtained from $R$ and $T$ data. The vertical broken lines in (e) and (f) indicate the photon energy of a femtosecond laser pulse used for the terahertz radiation experiments.

Table I. Obtained fitting parameters with $\varepsilon_\infty = 10.62$.

| Mode $i$ | $\omega_i$ (THz) | $\gamma_i$ (THz) | $f_i$ |
|---|---|---|---|
| 1 | 1.36 | 0.08 | 6.33 |
| 2 | 1.52 | 0.09 | 3.47 |
| 3 | 2.96 | 0.50 | 5.35 |
| 4 | 3.11 | 0.09 | 6.47 |
| 5 | 3.33 | 0.11 | 0.73 |
| 6 | 3.68 | 0.17 | 6.37 |
| 7 | 5.44 | 1.21 | 0.99 |
| 8 | 5.65 | 0.05 | 0.01 |
| 9 | 6.10 | 0.19 | 0.19 |
| 10 | 6.95 | 0.34 | 3.26 |
| 11 | 8.10 | 0.35 | 0.85 |
| 12 | 8.99 | 0.72 | 2.91 |